\documentclass[a4paper,twoside]{article}

\usepackage{epsfig}
\usepackage{subcaption}
\usepackage{calc}
\usepackage{amssymb}
\usepackage{amstext}
\usepackage{amsmath}
\usepackage{amsthm}
\usepackage{multicol}
\usepackage{pslatex}
\usepackage{apalike}
\usepackage{algorithm2e}
\usepackage[bottom]{footmisc}
\usepackage{SCITEPRESS}     

\usepackage{bm}
\usepackage{multirow}
\usepackage{makecell}
\usepackage{booktabs} 

\begin{document}

\title{EES-CND: Collaborative Neural Decision-Making for Drift-Aware Fault-Tolerant Edge-Cloud Service Placement}

\author{\authorname{Mohammadsadeq G. Herabad\sup{1}\orcidAuthor{0000-0002-2336-2077}, Javid Taheri\sup{1,2}\orcidAuthor{0000-0001-9194-010X}, Bestoun S. Ahmed\sup{1}\orcidAuthor{0000-0001-9051-7609} and Calin Curescu\sup{4}\orcidAuthor{0000-0001-5129-2718}}
\affiliation{\sup{1}Department of Mathematics and Computer Science , Karlstad University, Karlstad, Sweden}
\affiliation{\sup{2}School of Electronics, Electrical Engineering and Computer Science, Queen's University Belfast, Belfast, UK
}
\affiliation{\sup{4}Ericsson AB, Stockholm, Sweden}
\email{\{mohammadsadeq.garshasbi.herabad, bestoun, javid.taheri\}@kau.se, j.taheri@qub.ac.uk, calin.curescu@ericsson.com}
}


\abstract{The edge-cloud paradigm improves service delivery by orchestrating resources across edge nodes and cloud data centres. These environments consist of heterogeneous, interconnected computing nodes that cooperate to deliver continuous services. However, their scale and complexity increase vulnerability to failures from hardware malfunctions, software defects, and dynamic operating conditions. These failures can disrupt system configurations and service execution, leading to reduced reliability, performance degradation, and violations of service-level objectives. Ensuring service execution requires adaptive service placement strategies across edge-cloud resources. This study introduces a fault-tolerant service placement approach (Enhanced Evolution Strategy for Collaborative Neural Decision-making, EES-CND) for edge-cloud environments. The method employs collaborative decision-making, wherein multiple lightweight neural networks jointly infer redeployment strategies during failure events. To address the system dynamics and mitigate performance drift, adaptive models are updated online using an enhanced evolution strategy. Extensive simulations show that EES-CND effectively handles performance drift and significantly outperforms existing methods in service recovery time, response time, and reliability, achieving a 44.8\% reduction in fault-tolerance cost compared to standalone models.}

\onecolumn \maketitle \normalsize \setcounter{footnote}{0} \vfill

\section{\uppercase{Introduction}}
\label{sec:introduction}
Edge-cloud computing distributes computation across edge and cloud layers to support latency-sensitive applications. By bringing processing closer to users, it improves the Quality of Service (QoS) and reduces network congestion compared to cloud models. However, this paradigm introduces complexity, as distributed nodes operate under dynamic conditions and resource constraints, making them susceptible to failures caused by hardware faults, software defects, and workload fluctuations. These disruptions can lead to service-level agreement (SLA) violations, making fault-tolerant service placement critical in edge-cloud infrastructures \cite{mudassar2022adaptive}.

Designing an effective service redeployment strategy is challenging in failure-prone edge–cloud environments. Placement decisions must balance competing objectives (e.g., latency, resource utilisation, and reliability) under strict constraints, such as limited edge capacity and bandwidth. Optimising these objectives in dynamic and uncertain environments renders the problem highly complex and computationally intensive \cite{jia2023fault}. A key challenge arises from the nonstationarity of edge–cloud systems, in which failure patterns, resource availability, and workload characteristics evolve owing to hardware aging, software updates, fluctuating demand, and changing network conditions. These dynamics introduce data and concept drift, leading to performance degradation over time.

Conventional fault-tolerant service placement approaches mostly use static and offline optimisation models or heuristic rules based on fixed system assumptions. Although they are effective in short-term scenarios, they often lack the ability to adapt to changing failure patterns, workload fluctuations, and dynamic resource conditions. As drift occurs, these methods become misaligned with the operational environment, leading to suboptimal decisions and gradual performance degradation. In contrast, online learning-based approaches enable continuous system monitoring, iterative model updates, and adaptive placement decisions. Nevertheless, they introduce additional computational overhead, require stability control, and may face convergence challenges in highly non-stationary environments \cite{bhowmik2024dynamic}.

To address the aforementioned concerns, this study introduces a learning-driven fault-tolerant service placement algorithm based on Collaborative Neural Decision-making (CND) for dynamic edge–cloud environments. The proposed method formulates service placement as an online adaptive optimisation problem and employs a collaborative decision-making mechanism composed of multiple cooperating lightweight neural network models. Unlike existing approaches that depend on a single static or periodically retrained model, the proposed algorithm includes two parts: a set of pretrained models that capture stable system knowledge and a set of adaptive models that evolve online. The adaptive models are updated during system operation using an Enhanced Evolution Strategy (EES), enabling selective and efficient parameter adaptation in response to system dynamics. This design allows the algorithm to mitigate performance degradation and preserve decision stability. The main contributions of this study are as follows:

\begin{itemize}
    \item Formulating fault-tolerant service placement in edge–cloud environments as an online adaptive optimisation problem under dynamic failures, uncertainty, and performance drift.
    \item Proposing a learning-driven fault-tolerant placement approach using multiple cooperating neural networks in a unified architecture.
    \item Developing an enhanced evolution strategy to update adaptive models at runtime to mitigate performance drift.
    \item Conducting extensive experiments across diverse scenarios to demonstrate the effectiveness of the proposed approach.
\end{itemize}

The remainder of this paper is organised as follows. Section \ref{sec:relatedwork} reviews related work. Sections \ref{sysmodel}--\ref{ees-cnd} present the system model and the proposed approach. Section \ref{performance_metrics} describes the performance metrics used in this study. Sections \ref{setup}--\ref{conclusion} present the experimental results and conclusions, respectively.
\section{Related Work}\label{sec:relatedwork}
In the literature, edge-cloud fault tolerance is mostly categorised into reactive and proactive strategies. Many of them focus on generic faults without addressing drifts; this is one of the research gaps that is addressed in this study.

\subsection{Reactive Fault-tolerant Techniques}
Reactive methods are generally used in systems in which failures are difficult to predict or occur unpredictably. They use failure-detection mechanisms and trigger recovery actions. All computing systems are equipped with reactive recovery mechanisms, even when proactive strategies are used, because reactive handling is necessary when unexpected failures occur.

Various reactive fault tolerance strategies for edge computing have been proposed. \cite{mudassar2022adaptive} developed a hybrid checkpointing–replication mechanism, but it uses heuristic selection and assumes independent failures. It does not consider drift in system conditions over time. \cite{droob2023workrs} introduced an optimal Docker-based checkpointing model, though its evaluation was limited to small-scale Raspberry Pi clusters and does not address dynamic changes in large-scale environments. \cite{long2025fault} applied a Duelling DQN for offloading optimisation under Poisson-distributed server faults. This approach assumes stationary failure behaviour and does not consider online drift in system characteristics. Similarly, \cite{jia2023fault} utilised causal logging to reduce the microservice overhead; however, it was limited to single-node failures and did not address evolving system conditions or workload drift. \cite{ray2022prioritized} introduced a prioritised recovery framework based on stochastic multi-player games and probabilistic model checking. This method is computationally expensive and does not account for drift in system dynamics. \cite{bhowmik2024dynamic} proposed a fault-tolerant approach based on dynamic load balancing for parallel applications to redistribute tasks upon failures, but the method assumes homogeneous environments and does not handle time-varying system behaviour. \cite{javed2020edge} introduced an edge-based framework that uses subscription-based replication and container orchestration to handle network failures, but it was evaluated only on a homogeneous and small-scale setup and did not address online changes in system conditions.

Reactive methods are simple and resource-efficient because they do not require continuous monitoring, prediction models, or redundancy during normal operation. They are effective in handling unpredictable failures and are commonly used as baseline recovery mechanisms in most systems. However, because actions are triggered only after a failure occurs, they cause temporary service interruptions, latency spikes, or performance degradation during recovery.

\subsection{Proactive Fault-tolerant Techniques}
Proactive methods aim to anticipate and prevent failures before their occurrence. They utilise prediction models and system analytics to detect early signs of potential failures and undertake preventive actions. These approaches are used in environments where failure patterns can be estimated or predicted, allowing the system to maintain service continuity and reduce the likelihood of disruptions.

\cite{tuli2022pregan} proposed a proactive GAN-based fault-tolerance framework for edge–cloud systems that predicts preemptive migration using graph attention and prototype-based fault classification. It assumes relatively stable system conditions without addressing the drift in system dynamics. In another study, \cite{tuli2023pregan} developed a semi-supervised variant that integrates graph attention, temporal modelling, and transformer-based fine-tuning to improve fault detection. It further increases the architectural complexity and overhead, but does not consider evolving system conditions or drift. \cite{sahu2025adaptive} presented a proactive method for Digital Twins using a hybrid Genetic–PSO algorithm to optimise recovery metrics. It incurs considerable computational overhead because of metaheuristic optimisation, yet does not address the non-stationary system behaviour. \cite{liu2023service} proposed a two-phase method combining time-series fault prediction with GA-based container migration to enhance edge service reliability. It assumes homogeneous devices, depends on labelled historical failure data, without considering drift or changing system conditions over time. \cite{arun2019ezbft} proposed a proactive replication-based protocol in which any replica can act as a leader, using dependency tracking and speculative execution to reduce communication steps and latency. It assumes synchronous and homogeneous networks without considering the evolving operational conditions. \cite{theodoropoulos2022automated} used multi-step deep learning to manage workload-induced congestion instead of node failures. They also did not address drift in the system behaviour. \cite{tuli2023deepft} proposed a self-supervised surrogate model to predict overloads, though it introduces high computational complexity through online fine-tuning and does not consider system drift. \cite{long2022novel} utilised a Deep Q-Network for Primary–Backup replication in IoT environments, but it uses predefined failure mechanisms and stationary assumptions that may lack flexibility in dynamic environments.

The main advantage of proactive methods is that they can reduce service interruptions and improve system reliability by acting before failures occur. However, they require continuous monitoring, prediction mechanisms, and redundancy, which introduce additional computational, communication, and resource overheads. Moreover, their effectiveness depends on the accuracy of failure prediction, and incorrect predictions may lead to unnecessary recovery actions.

\subsection{Research Gaps}
Despite the significant body of work on both reactive and proactive fault-tolerance mechanisms for edge–cloud computing, system drift and evolving operational conditions have not been sufficiently addressed in the literature. Many existing approaches consider the assumptions of stationary system behaviour and fixed resource and workload patterns. However, real-world edge–cloud environments are dynamic and non-stationary, in which resource availability, network conditions, workload characteristics, and service interactions may change over time. Under these conditions, models trained offline or designed for static environments would gradually lose their efficacy, and drift-aware and online-adaptive fault-tolerance mechanisms remain an open research problem. This motivates the need for approaches that can adapt to continuously evolving system conditions.
\section{\uppercase{System Model}}\label{sysmodel}
\begin{figure}
    \centering
    \includegraphics[width=0.8\linewidth]{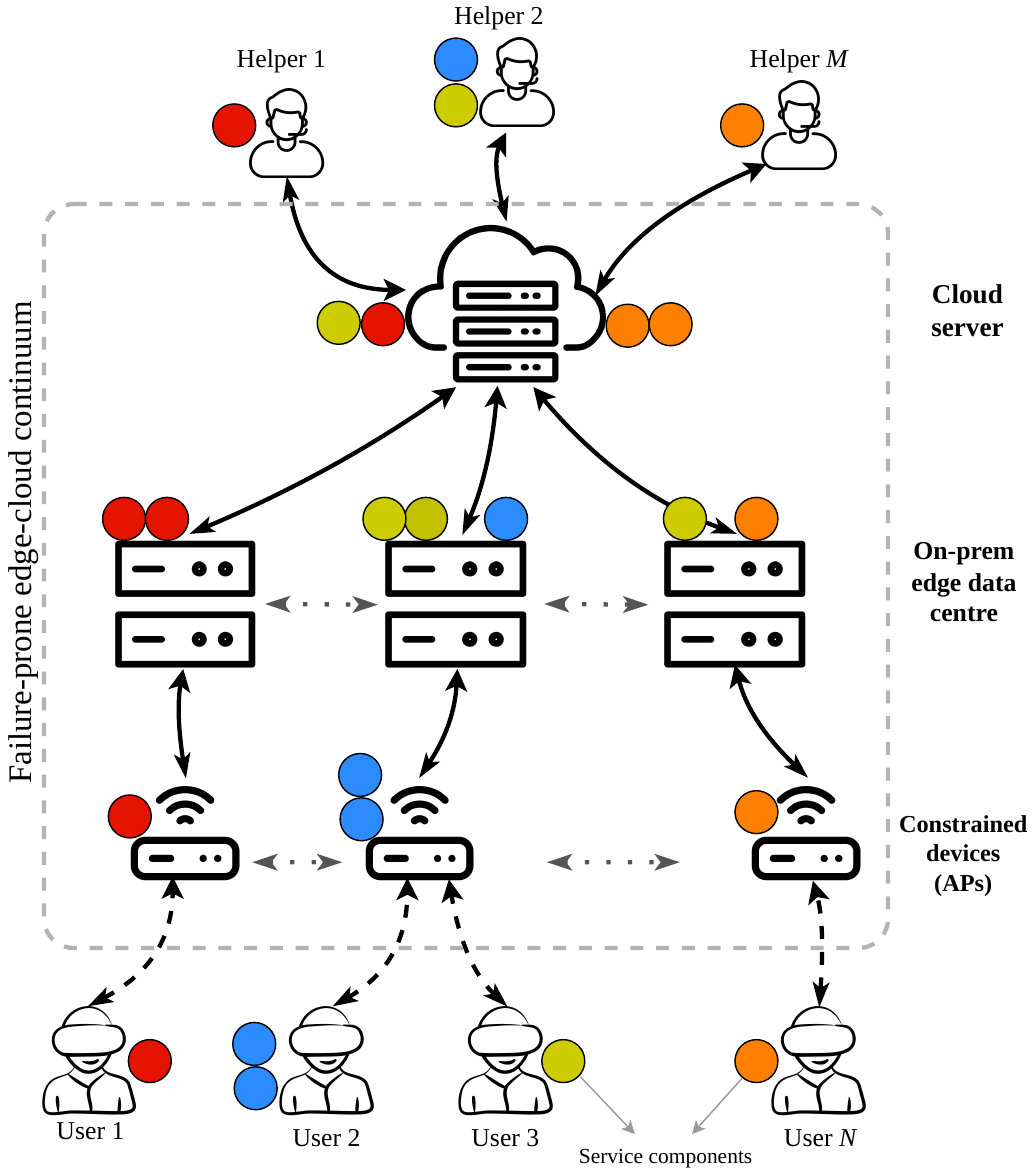}
    \caption{Failure-Prone Edge–Cloud Continuum}
    \label{fig:infra}
\end{figure}

This study considers a failure-prone edge–cloud architecture modelled as a distributed system of heterogeneous computing nodes. Although this system can be applied to various edge–cloud computing scenarios, we consider a use case to demonstrate our solution. This use case was developed with our industrial partner (Ericsson) as a representative scenario (Figure \ref{fig:infra}). In our use case, users interact with remote helpers for repair tasks through service components across the edge–cloud continuum. Table \ref{tab:notations} presents the notations used in this section.

\subsection{Computing Nodes}
The infrastructure comprises three logical layers: (i) access points (APs), (ii) edge servers, and (iii) a cloud layer. $\mathcal{CN}(t) = \lbrace CN_1(t), CN_2(t), ..., CN_K(t) \rbrace$ defines the set of computing nodes at time $t$. Each computing node is characterised by a multidimensional vector, ${CN}_k(t) = \langle CC_k(t), MC_k(t), DC_k(t), RS_k(t)\rangle$, where $CC_k(t)$, $MC_k(t)$, and $DC_k(t)$ denote the computational, memory, and disk capacities of node $k$, respectively, and $RS_k(t) \in (0,1]$ shows its reliability score. All computing nodes are assumed to be vulnerable to unexpected failures, including those with high reliability \cite{javed2020edge}.

\subsection{User and Helper Nodes}
An edge-cloud scenario is considered in which users collaborate with remote helpers for maintenance and repair tasks. $\mathcal{U} = \{u_1, \dots, u_N\}$ and $\mathcal{H} = \{h_1, \dots, h_M\}$ denote the sets of user and helper devices, respectively. Each user/helper node is defined by a resource tuple $\langle CC_i, MC_i, DC_i, RS_i \rangle$, representing the computing, memory, disk capacities, and reliability score. An active session at time $t$ is denoted by $p_{n,m}(t)$ which represents a real-time connection between user $u_n$ and helper $h_m$.

\subsection{Network Communications}
The available bandwidth and transmission delay are observable or estimable for any directed link at time $t$ \cite{cozzolino2022nimbus}. We model the communication links as a weighted directed graph, $\mathcal{G} = (\mathcal{K}, \mathcal{E})$, where
$\mathcal{E} \subseteq \mathcal{K} \times \mathcal{K}$ represents the set of communication links between the nodes. $\mathcal{K}$ denotes the set of all network entities.
A single communication characteristic matrix is defined, $\mathcal{L}(t) = [\ell_{ij}(t)]$, where each element is a two-dimensional vector: $\ell_{ij}(t) = \langle b_{ij}(t), d_{ij}(t) \rangle$, with $b_{ij}(t)$ denoting the available bandwidth  and $d_{ij}(t)$ representing the observed transmission delay  from entity $i$ to entity $j$ at time $t$. This representation reconciles structural node indexing with dynamic link properties (Equation \eqref{communication}).

\begin{equation}\label{communication}
\ell_{ij}(t) =
\begin{cases}
(b_{ij}(t), d_{ij}(t), & \text{if } (i,j) \in \mathcal{E}, \\
(0, +\infty), & \text{otherwise}.
\end{cases}
\end{equation}

\subsection{Applications (Services)}
The system hosts a set of services $\mathcal{S} = \{S_1, \dots, S_X\}$, where each service $S_x$ is modelled as a Directed Acyclic Graph (DAG), $\mathcal{G}_x = (\mathcal{C}_x, \mathcal{E}_x)$. $\mathcal{C}_x$ denotes the service components, and $\mathcal{E}_x$ represents the data-flow dependencies. Each component $SC_y^x \in \mathcal{C}_x$ is available in multiple versions $SC_y^x = \{SC_{y,1}^x, \dots, SC_{y,V}^x \}$, reflecting different implementations. Each version is characterised by a resource demand profile (CPU, memory, and storage requirements) and a reliability score.

\begin{table}
\centering
\caption{Summary of system model notations}
\label{tab:notations}

\setlength{\tabcolsep}{3pt}
\renewcommand{\arraystretch}{0.9}

\resizebox{0.99\linewidth}{!}{%
\begin{tabular}{ll}
\hline
\textbf{Notation} & \textbf{Description} \\
\hline
$CN_k$ & $k^{th}$ computing node\\
$\mathcal{CN}(t)$ & Set of computing nodes and characteristics at $t$\\
$\mathcal{U}$ & User nodes $\{u_1,\dots,u_N\}$ \\
$\mathcal{H}$ & Helper nodes $\{h_1,\dots,h_M\}$ \\
$u_n/h_m$ & $n^{th}$ user / $m^{th}$ helper node \\
$K/N/M$ & Number of nodes/users/helpers \\
$\mathcal{S}$ & Services $\{S_1,\dots,S_X\}$ \\
$S_x$ & Service of user–helper pair $x$ \\
$SC_y^x$ & $y^{th}$ component of $S_x$ \\
$SC_{y,v}^x$ & Version $v$ of component $y$ in $S_x$ \\
$\mathcal{C}_x$ & Components of service $S_x$ \\
$X/Y/V$ & Number of services/components/versions \\
$\mathcal{L}(t)$ & Communication link matrix \\
$b_{ij}(t)$ & Bandwidth between nodes $i$ and $j$ \\
$d_{ij}(t)$ & One-way delay between nodes $i$ and $j$ \\
$\ell_{ij}(t)$ & Link characteristics $b_{ij}(t)$ and $d_{ij}(t)$ \\
\hline
\end{tabular}
}
\end{table}

\subsection{\uppercase{Drift Model}}
Drift refers to the gradual change in edge-cloud system characteristics over time, wherein the main parameters, such as node resources, network conditions, and service demands, change gradually and/or continuously \cite{mehmood2024lstmdd}. These changes modify the relationships between system components and alter the optimal service deployment configuration \cite{sunaga2024sequential}. To represent this behaviour, we applied a multi-level incremental drift to the system model \cite{wang2022concept}, allowing the parameters to change gradually. At the node level, drift affects the computational and memory capacities and their interdependencies. At the network level, it alters the communication characteristics, including latency, bandwidth, and their statistical correlations. Service-level drift is also incorporated by varying the resource demands of service components over time. In this online setting, future characteristics are unknown, and redeployment decisions must be made based on the currently observable system state.
\section{\uppercase{Proposed Approach}}\label{ees-cnd}

\subsection{Monitoring and Fault Detection}
In the proposed approach, a monitoring system tracks the health status and performance of computing nodes and deployed service components. Each node and component periodically sends lightweight heartbeat signals to a centralised controller, serving as liveness indicators. These heartbeats are generated at fixed intervals (e.g., every 10 s in Kubernetes-based systems \cite{kubernetes_pod_lifecycle}). During the fault detection phase, the absence of heartbeat messages within a predefined timeout interval indicates failure. Once a fault is detected, the controller determines the scope of the failure by identifying all affected entities. Subsequently, a consolidated list of failed components is generated. This list serves as the recovery input for the proposed placement algorithm. The redeployment algorithm then computes a recovery strategy for the failed component. 

\subsection{EES-CND (our proposed solution)}
\subsubsection{Input and Output}
The proposed approach (enhanced evolution strategy for collaborative neural decision-making, EES-CND) is triggered upon failure and executed independently for each computing node. It takes as input a high-dimensional representation of the system state at time $t$, defined as $s(t) = \left\{ CN_k(t), L(t), C_x(t), RE_{CN_k}(t) \right\}$. $CN_k(t)$ denotes the resource characteristics of candidate node $k$, $L(t)$ represents the network state (bandwidth and delay), $C_x(t)$ captures the demand for the failed component $x$, and $RE_{CN_k}(t)$ is the reconfiguration cost of deploying the component at node $k$. The EES-CND mechanism is executed for each candidate node to estimate its suitability for hosting failed components. For a given failure interval $\tau$, the cost is computed for every candidate node, reflecting the impact of assigning the failed components under the current system state. The redeployment target is then selected via minimisation (i.e., choosing the node with the minimum deployment cost).

\subsubsection{Architecture and Operation}
\begin{figure}
    \includegraphics[width=0.9 \linewidth]{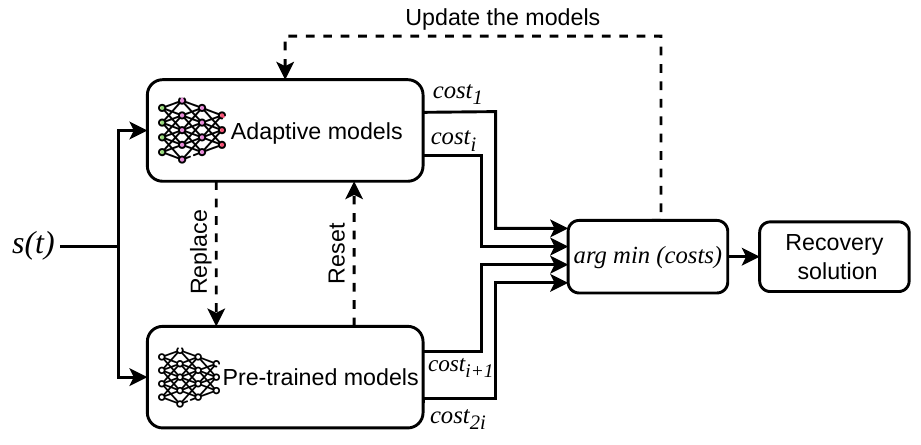}
    \caption{Collaborative neural decision-making approach}
    \label{fig:CND}
\end{figure}

The EES-CND architecture follows a two-tier design (a pretrained model group and an adaptive model group), both operating on the same system state input. Figure \ref{fig:CND} illustrates the high-level operational workflow and architecture of the EES-CND. The objective is to generate a recovery decision by evaluating candidate nodes and selecting the one that minimises the cost (node suitability). The pretrained models consist of multiple neural networks that are trained offline to capture promising placement patterns or algorithms. Their parameters remain fixed during runtime, providing stable and historically informed decisions. The adaptive models contain the same number of networks, where each pretrained model has a structurally identical adaptive counterpart (i.e., the same layers and neurones). Unlike pretrained models, adaptive models are randomly initialised and updated online to adjust to evolving system conditions (drifts). We empirically considered 12 models in each group: six pretrained models and six corresponding adaptive counterparts. At each failure interval, every neural network in both groups outputs the cost for its selected candidate node. These outputs are aggregated, and a global minimisation step selects the node with the lowest cost as the final recovery decision. This competitive selection enables cross-model comparisons and ensures that the redeployment target is determined by the best-performing model at that interval.

\begin{figure}
    \includegraphics[width=0.9 \linewidth]{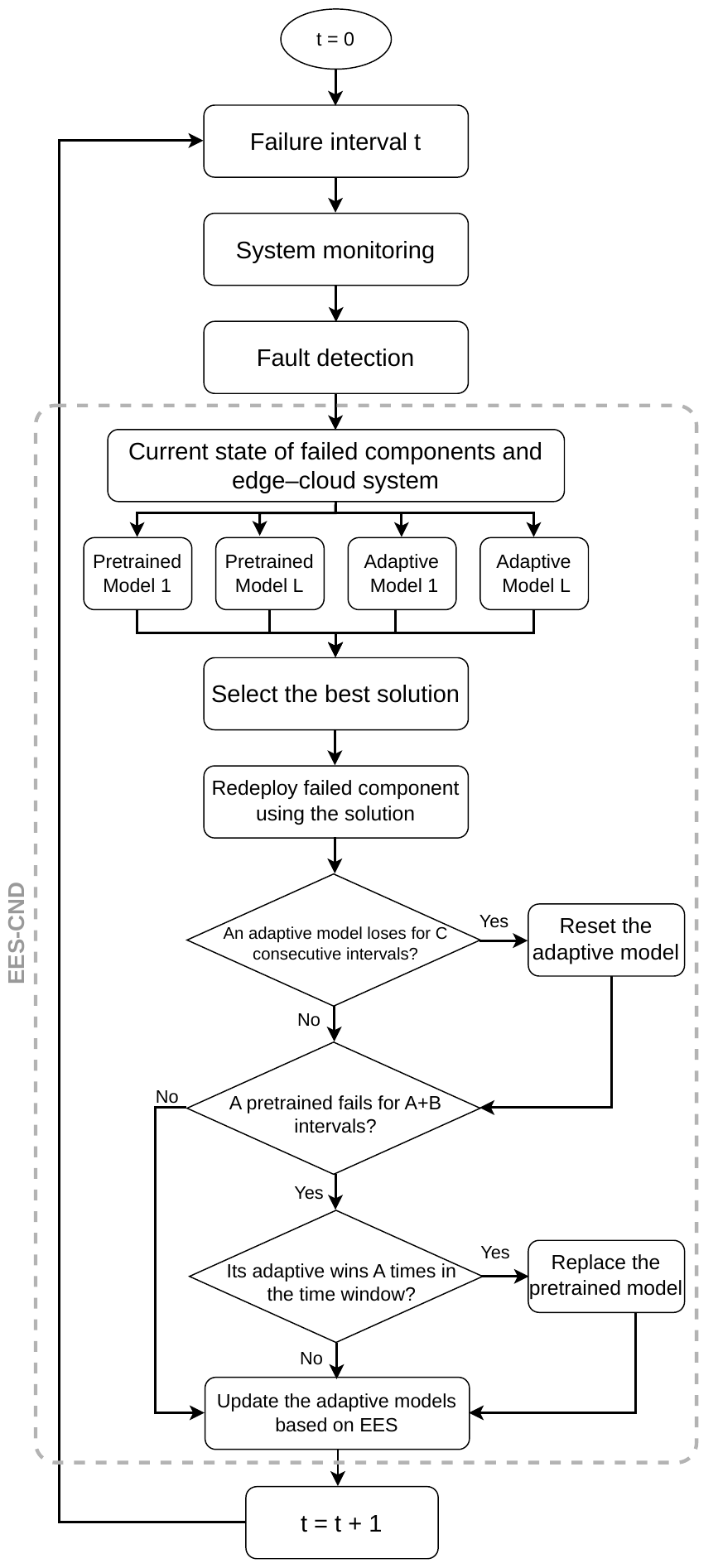}
    \caption{EES-CND flowchart}
    \label{fig:flowchart}
\end{figure}

The adaptive models are updated at each failure interval. To avoid stagnation and preserve diversity, a reset mechanism is introduced: if an adaptive model does not win the global selection for $C$ consecutive intervals, its parameters are reinitialised to escape suboptimal regions. A controlled replacement mechanism is also applied between model pairs: if a pre-trained model fails to outperform its adaptive counterpart for ${A} + {B}$ consecutive intervals, while the adaptive model wins at least ${A}$ times within that time window, the pre-trained model is replaced. This allows successful adaptive models to update the base model pool and transfer the learned improvements into the stable group. Figure \ref{fig:flowchart} illustrates the flowchart of the proposed approach.

\subsubsection{Enhanced Evolution Strategy (EES)}
Evolution strategy (ES) \cite{salimans2017evolution} is a gradient-free optimisation method that searches for better solutions using stochastic parameter perturbations and fitness-based updates. The proposed approach extends the standard ES by introducing parameter-wise adaptive exploration instead of a fixed isotropic exploration rate for updating the adaptive models. Unlike conventional ES, where all parameters share the same perturbation scale (which may lead to an inefficient search when different parameters have different sensitivities), the proposed EES adjusts the exploration scale of each parameter based on its estimated importance during optimisation. In the proposed approach, the adaptive models are updated online at each failure interval using EES.

As shown in Algorithm~\ref{alg:enhanced_es}, EES-CND first generates symmetric candidate solutions around the current parameters using stochastic perturbations and antithetic sampling and evaluates them using the fault tolerance cost function defined in Equation~\eqref{obj} (lines 4--5). The obtained costs are then converted into utility values using rank-based normalisation to improve robustness against scale variations and outliers (line 7). Based on these utilities, a stochastic gradient estimate is computed to determine the search direction in the parameter space (line 8), and the model parameters are subsequently updated (line 9). Unlike the standard ES, the EES introduces an additional adaptive mechanism that estimates the importance of each parameter based on the magnitude of the gradient components (line 10). These importance scores are then used to update the exploration vector, thereby allowing each parameter to adapt its perturbation scale during optimisation (line 11). For numerical stability, the exploration values are bounded within predefined limits when such bounds are specified (lines 12--14). Table \ref{tab:notations3} summarises the EES-related notations.

In the considered online setting, only a single sample is available per failure interval, making gradient-based methods, such as backpropagation, unreliable owing to unstable gradient estimation. In contrast, the EES operates gradient-free through stochastic perturbations and direct cost evaluation, thereby enabling stable online adaptation at each failure interval.

\begin{table}
\centering
\caption{Summary of EES notations}
\label{tab:notations3}

\setlength{\tabcolsep}{3pt}
\renewcommand{\arraystretch}{0.9}

\resizebox{0.99\linewidth}{!}{%
\begin{tabular}{ll}
\hline
\textbf{Notation} & \textbf{Description} \\
\hline
$\theta_t$ & Model parameters at time $t$\\
$\sigma_t$ & Exploration vector at time $t$\\
$\epsilon_i$ & Perturbation at interval $i$\\
$\theta_i^{\pm}$ & Symmetric parameters generated from $\theta_t$\\
$c_i^{\pm}$ & Fault-tolerance cost of $\theta_i^{\pm}$\\
$u_i^{+},u_i^{-}$ & Utility values\\
$g_t$ & Exploration direction in parameter space\\
$I_{t,j}$ & Importance of parameter $j$\\
$\alpha$ & Learning rate\\
$\mathcal{K}$ & Number of samples\\
$\eta_{\sigma}$ & Exploration learning rate\\
\hline
\end{tabular}
}
\end{table}

\begin{algorithm}[t]
\caption{Enhanced Evolution Strategies (EES)}
\label{alg:enhanced_es}
\LinesNumbered
\small
\KwData{${\theta}_0$, ${\sigma}_0$, $\alpha$, $\eta_{\sigma}$, $K$, $\sigma_{\min}, \sigma_{\max}$}
\KwResult{Updated ${\theta}$ and ${\sigma}$}

Initialise $\tau \leftarrow 0$\;

\While{stopping criterion not met}{
    
    \For{$i \leftarrow 1$ \KwTo $K$}{
        Sample ${\epsilon}_i \sim \mathcal{N}(\mathbf{0},\mathbf{I})$
        
        Evaluate $c_i^{\pm} \leftarrow 
        \mathrm{cost}({\theta}_t \pm {\sigma}_t \odot {\epsilon}_i)$\;
    }

    Rank-normalise $\{c_i^{+},c_i^{-}\}$ to utilities $\{u_i^{+},u_i^{-}\}$\;

    ${g}_t \leftarrow 
    \frac{1}{K}\sum_{i=1}^{K}
    (u_i^{+}-u_i^{-})({\epsilon}_i \oslash {\sigma}_t)$

    ${\theta}_{t+1} \leftarrow {\theta}_t + \alpha {g}_t$

    $\tilde{{I}}_t \leftarrow 
    \frac{|{g}_t|}{\mathrm{mean}(|{g}_t|)}$\;

    ${\sigma}_{t+1} \leftarrow 
    {\sigma}_t \odot 
    \exp\!\big(\eta_\sigma(\tilde{{I}}_t - \mathbf{1})\big)$\;

    \If{bounds specified}{
        ${\sigma}_{t+1} \leftarrow 
        \mathrm{clip}({\sigma}_{t+1}, 
        \sigma_{\min}, \sigma_{\max})$\;
    }

    $\tau \leftarrow \tau + 1$\;
}
\end{algorithm}

\subsubsection{Pretrained Models}
The pretrained models are trained through an offline procedure based on the approach introduced by \cite{garshasbi2025lightweight}. The model parameters are trained using a Genetic Algorithm (GA) that directly minimises the fault-tolerance cost. The neural network parameters are encoded as candidate solutions in the GA population. An initial population is randomly generated to ensure diversity, and each candidate is evaluated using the defined objective. The population is then iteratively refined through selection, crossover, and mutation operators. The GA is adopted for offline pre-training to exploit its global search capabilities in complex and gradient-free parameter spaces, because the GA's high computational overhead makes it unsuitable for online use. Nevertheless, we use a GA (offline ) to learn robust initial pre-training models and an EES for online adjustment, making it a lightweight mechanism for real-time (online) adaptation.
\section{\uppercase{Performance metrics}}\label{performance_metrics}

\subsection{Recovery Time}
Two types of failures exist \cite{arun2019ezbft}: computing node and service component failures. Node failure causes the hosted components to fail, whereas service components may fail independently. Service recovery ($RE_S$) involves fault detection, decision making, and reconfiguration. Fault detection occurs via heartbeat monitoring, in which components periodically send heartbeats to the monitoring controller. If the interval exceeds the threshold $t_{th}$, it is marked as \textit{Not Ready}. The decision phase runs placement optimisation to determine the new deployment locations or component versions.

For a failed service component in set $\mathcal{F}$, Equation~\eqref{reconf} computes the reconfiguration time, where $Size_{y,v}^x$ is the container image size, $a_{y,v}^x$ is the instantiation time, and $\mathbb{I}_{y,v}^x\in\{0,1\}$ indicates whether the image is already available on the target node. $BW_l(t)$ denotes the available bandwidth of link $ l $ at time $t$. Equation~\eqref{maxReconf} provides the overall reconfiguration time, which is dominated by the slowest redeployed component.

\begin{equation}\label{reconf}
C_{SC_y^x}= a_{y,v}^x + (1-\mathbb{I}_{y,v}^x)\,\frac{Size_{y,v}^x}{BW_l(t)}
\end{equation}

\begin{equation}\label{maxReconf}
C =\max_{SC_y^x\in\mathcal{F}} C_{SC_y^x}
\end{equation}

\subsection{Response Time}
Equation \eqref{td} \cite{cozzolino2022nimbus} computes transmission delay, where $DS_{SC_{y,v}^x}$ is the transmitted data size of the service components ${SC_{y,v}^x}$, $b_{ij}(t)$ is the available bandwidth between nodes $i$ and $j$ to transfer the data, and $d_{ij}(t)$ is the propagation delay between nodes $i$ and $j$ at time $t$.

\begin{equation}\label{td}
TD_{SC_{y,v}^x} = \frac{DS_{SC_{y,v}^x}(t)}{b_{ij}(t)} + d_{ij}(t)
\end{equation}

Equation \eqref{et} \cite{yeganeh2023novel} calculates the execution time of a service component, where $CR_{SC_{y,v}^x}$ denotes the computational requirement of $SC_{y,v}^x$ and $CC_k(t)$ denotes the computational capacity of node $k$ at time $t$.

\begin{equation}\label{et}
ET_{SC_{y,v}^x}
=
\frac{CR_{SC_{y,v}^x}(t)}{CC_k(t)}
\end{equation}

Provider delay $PD_{SC_{y,v}^x}$ captures the latency from the service provider (e.g., virtualisation or queuing) and is estimated using passive measurements. The encoding/decoding delay ($CD_{SC_{y,v}^x}$) represents the time required to compress and decompress video streams. These delays are provided by the codec provider. Equation \eqref{restime} defines the response time of the service component $SC_{y,v}^x$ deployed on the computing node. Equations \eqref{ssrt} and \eqref{tsrt} calculate the service-level response time and total response time across all services, respectively \cite{herabad2025psoga}.

\begin{align} \label{restime}
RT_{SC_{y,v}^x} &= TD_{SC_{y,v}^x} + ET_{SC_{y,v}^x}
 + PD_{SC_{y,v}^x} + CD_{SC_{y,v}^x}
\end{align}

\begin{equation}\label{ssrt}
RT_{S_x}
=
\sum_{y=1}^{Y}
RT_{SC_{y,v}^x}
\end{equation}

\begin{equation}\label{tsrt}
RT_S
=
\sum_{x=1}^{X}
RT_{S_x}
\end{equation}

\subsection{Service Reliability}
For a component version or node with a failure rate $\lambda$, the reliability over the time horizon $t_h$ is given by $e^{-\lambda t_h}$ \cite{mudassar2022adaptive}. The reliability of $SC_{y,v}^x$ is computed as $e^{-\lambda_{y,v}^{x} \cdot t_h}$, and the reliability of $CN_k$ is computed as $e^{-\lambda_k \cdot t_h}$.

The system is modelled as a component chain, in which any component failure causes a service outage. Therefore, Equation~\eqref{srel} estimates the software reliability, and Equation~\eqref{hrel} computes the hardware reliability for node failures, respectively \cite{mudassar2022adaptive}.
\begin{equation}\label{srel}
R_{S_x}^{\text{soft}}
=
\exp\left(
- t_h \sum_{SC_{y,v}^{x} \in \mathcal{C}_x}
\lambda_{y,v}^{x}
\right)
\end{equation}

\begin{equation}\label{hrel}
R_{S_x}^{\text{hard}}
=
\exp\left(
- t_h \sum_{CN_k \in \mathcal{CN}}
\lambda_k
\right)
\end{equation}

Considering the interdependence between software and hardware failures, Equation \eqref{orel} calculates the overall service reliability of $S_x$. Equation \eqref{averel} computes the average reliability of all services.

\begin{equation}\label{orel}
R_{S_x}
=
R_{S_x}^{\text{soft}}
\cdot
R_{S_x}^{\text{hard}}
\end{equation}

\begin{equation}\label{averel}
RS_{{S}} = \frac{1}{X} \sum_{x=1}^{X} R_{S_x}
\end{equation}

\begin{table}
\centering
\caption{Summary of performance metric notations}
\label{tab:notations2}

\setlength{\tabcolsep}{4pt}
\renewcommand{\arraystretch}{0.9}

\resizebox{0.99\linewidth}{!}{%
\begin{tabular}{ll}
\hline
\textbf{Notation} & \textbf{Description} \\
\hline
$RE_S$ & Service recovery time\\
$C_{SC_y^x}$ & Reconfiguration time of $SC_y^x$\\
$BW_l(t)$ & Link bandwidth for image download at $t$\\
$a_{y,v}^x$ & Container launch time of $SC_y^x$\\
$Size_{y,v}^x$ & Image size of $SC_y^x$\\
$\mathcal{F}$ & Set of failed components\\
$DS_{y,v}^x$ & Data size transferred by $SC_{y,v}^x$ \\
$CR_{SC_{y,v}^x}$ & Compute requirement of $SC_{y,v}^x$\\
$CC_k(t)$ & Node $k$ compute capacity at $t$ \\
$TD_{SC_{y,v}^x}$ & Transmission delay of $SC_{y,v}^x$\\
$ET_{SC_{y,v}^x}$ & Execution time of $SC_{y,v}^x$\\
$PD_{SC_{y,v}^x}$ & Provider delay of $SC_{y,v}^x$\\
$CD_{SC_{y,v}^x}$ & Encoding/decoding delay of $SC_{y,v}^x$\\
$RT_{SC_{y,v}^x}$ & Response time of $SC_{y,v}^x$\\
$RT_{S_x}$ & Response time of $S_x$\\
$R_{S_x}^{\text{soft}}$,  $R_{S_x}^{\text{hard}}$ & Software and hardware reliability of $S_x$\\
$R_{S_x}$ & Overall reliability of $S_x$\\
$\lambda$ & Failure rate\\
$res(CN_k)$ & Resource capacity of $CN_k$\\
$res(SC_{y,v}^x)$ & Resource requirement of $SC_{y,v}^x$\\
$u^{SC}_n$ & Components belonging to $u_n$\\
$h^{SC}_m$ & Components belonging to $h_m$\\
$w_1,w_2,w_3$ & Objective weights \\
\hline
\end{tabular}
}
\end{table}

\subsection{Objective Function}
The objective is to minimise the fault tolerance cost. Equation \eqref{obj} computes the fault-tolerance cost by combining the normalised metrics using weighted-sum scalarization. $\widetilde{RT}_S$, $\widetilde{RS}_S$, and $\widetilde{RE}_S$ represent the normalised response time, reliability, and reconfiguration overhead, respectively. Equal weights ($w_i = 0.33$) are used by default but can be adjusted based on the system priorities. To establish a minimisation objective, the reliability metric is transformed into its complement, $1 - \widetilde{RS}_S$. Therefore, the redeployment problem is formulated as a constrained optimisation problem (Equation \eqref{pobj}).

\begin{equation}\label{obj}
FT_{cost} = w_1 \cdot \widetilde{{RT}_S} + w_2 \cdot (1 - \widetilde{{RS}_S}) + w_3 \cdot \widetilde{{RE}_S}
\end{equation}

\begin{equation}\label{pobj}
\min \; FT_{\text{cost}}
\end{equation}

subject to:
\begin{equation}\label{c1}
\sum_{v \in V} SC_{y,v}^x = 1, \quad \forall y \in Y, \quad SC_{y,v}^x \in \{0,1\}
\end{equation}
\begin{equation}\label{c2}
\sum_{x,y\in X,Y} res(SC_{y,v}^x) \:<\: res(CN_k), \quad \forall k \in K
\end{equation}
\begin{equation}\label{c3}
SC_{y,v}^x \in u^{SC}_n, \quad \forall x \in X, \forall v \in V
\end{equation}
\begin{equation}\label{c4}
SC_{y,v}^x \in h^{SC}_m, \quad \forall x \in X, \forall v \in V
\end{equation}

Constraint~\eqref{c1} ensures that each service component is deployed with one version. Constraint~\eqref{c2} limits the total resource demand on node $k$ to its capacity. Constraints~\eqref{c3} and \eqref{c4} restrict the deployment to user or helper nodes associated with the service. Table \ref{tab:notations2} summarises the performance metric notations.
\section{\uppercase{Experimental Setup}}\label{setup}
The evaluation was conducted using the simulator introduced in \cite{garshasbi2024optimizing}, which reflects real-world edge–cloud environments, particularly for remote repair and maintenance use cases. It models characteristics such as multi-version components, user–helper interactions, and multi-tier edge–cloud architectures (features rarely supported in existing simulators). For reproducibility, all simulation artefacts and implementation details are available on GitHub \cite{simulator}.

\begin{figure*}
    \centering
    \includegraphics[width=0.97\linewidth]{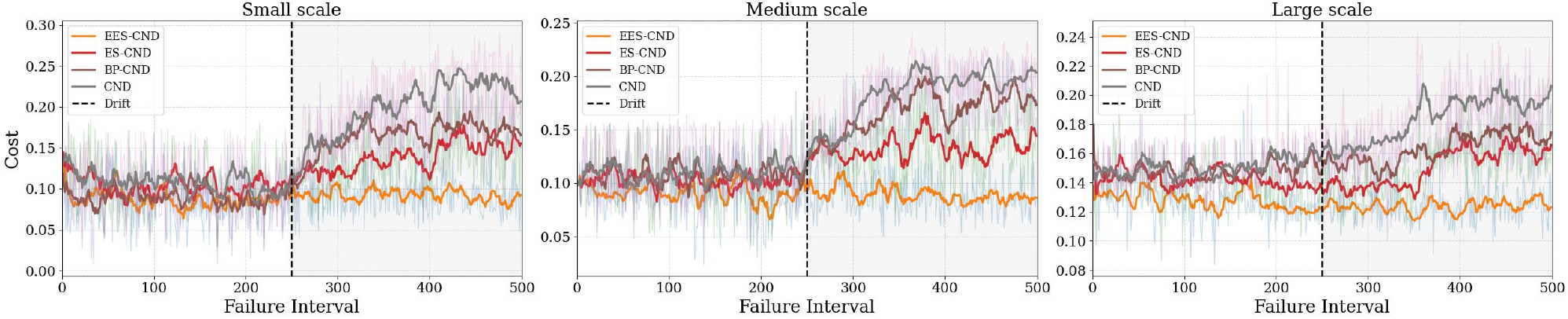}
    \caption{Overall cost achieved by the algorithms (pale lines: actual data; plain lines: 10-point moving average).}
    \label{fig:cost}
\end{figure*}

\subsection{Problem Instances}
The evaluation employs two distinct problem sets: (i) an offline training set comprising small, medium, and large-scale instances to capture varying system complexities (see Table \ref{tab:evaluation_scales}) and (ii) an online test set of a comparable scale but differing configurations to evaluate generalisation and prevent overfitting. To simulate non-stationary edge-cloud environments, the test instances incorporate multilevel incremental drift, which affects node resources, network conditions, and service demands, introduced to model continuous system evolution. Instance characteristics, including network properties (bandwidth and delay), node attributes (CPU, memory, and storage), and service requirements (computational demand, memory need, and data size), are aligned with the specifications in \cite{simulator}.

\begin{table}
\centering
\caption{Evaluation scales}
\label{tab:evaluation_scales}
\resizebox{0.99\linewidth}{!}{%
\begin{tabular}{lccc}
\toprule
\textbf{Specifications} & \textbf{Small-scale} & \textbf{Medium-scale} & \textbf{Large-scale} \\
\toprule
No.\ of user-helper pairs   & 25  & 50  & 100 \\
\midrule
No.\ of APs                 & 25  & 50  & 100 \\
\midrule
No.\ of edge servers        & 12  & 25  & 50  \\
\midrule
No.\ of cloud nodes         & 6   & 12  & 25  \\
\midrule
No.\ of service comp./vers. & 6/7 & 7/7 & 8/7 \\
\bottomrule
\end{tabular}%
}
\end{table}

\subsection{Assumptions and Benchmarks}
The system operates in discrete time intervals \cite{jagannathan2025towards}. At each interval, a subset of entities fails (either randomly or based on reliability scores). Specifically, we assume a failure ratio of 1\%--5\% \cite{mudassar2022adaptive} in each interval for both computing nodes (triggering cascading component failures) and individual service components (due to software-related faults). We also assume that edge--cloud resources are pre-provisioned and available prior to operation. Failures occur during execution without prior reconfiguration. It is assumed that a failure at interval $\tau$ is detected and recovered in $\tau+1$. The system is equipped with reliable monitoring to ensure accurate status and utilisation measurements \cite{mudassar2022adaptive}.

To evaluate the proposed approach (EES-CND), we compared it with three comparative approaches with the same goals: (i) BP-CND, which uses gradient-based backpropagation for model updates; (ii) ES-CND, which employs standard evolution strategies for model updates; and (iii) CND, which lacks an adaptive layer and uses only pretrained models. In addition, we tested a single-model configuration to isolate the impact of the collaborative mechanism. The hyperparameters and model architectures of the EES-CND, obtained empirically, are listed in Table \ref{tab:parameters}.

\begin{figure*}
    \centering
    \includegraphics[width=0.97\linewidth]{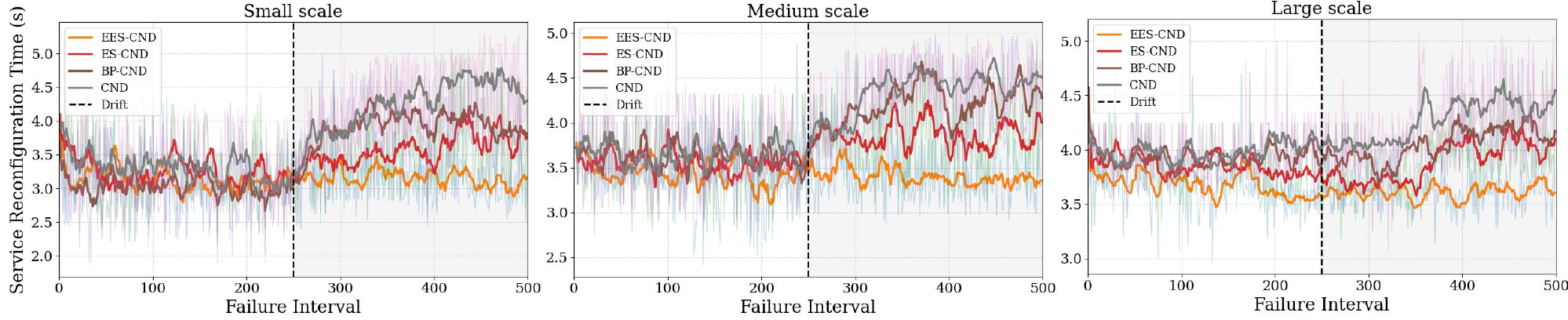}
    \caption{Service reconfiguration time (pale lines: actual data; plain lines: 10-point moving average)}
    \label{fig:reconf}
\end{figure*}

\begin{table}

\caption{EES-CND architecture and configurations}
\label{tab:parameters}
\resizebox{0.99\linewidth}{!}{%
\begin{tabular}{ll}

\toprule
\textbf{Hyperparameter} & \textbf{Value} \\
\toprule
Learning and exploration rate ($\alpha$ , $\eta_{\sigma}$)  &  0.1, 0.05  \\
\midrule
Sample size ($\mathcal{K}$)  &  10   \\
\midrule
Initial exploration ($\sigma_0$)  &   0.25   \\
\midrule
Exploration bounds ($\sigma_{min}$ and $\sigma_{max}$) &  1 and 1e-4 \\
\midrule
{Reset threshold of adaptive models (${C}$)}  & 50 intervals \\
\midrule
Replacement parameters (${A}$, ${B}$)  &  50, 25 intervals\\
\midrule
Number of the models in each group  & 6  \\
\midrule
\multirow{6}{*}{ Model architecture (layers and neurons)} & \textit{Model 1 (M1):} Input-5-output\\
                                    & \textit{Model 2 (M2):} Input-15-output \\
                                    & \textit{Model 3 (M3):} Input-8-4-output\\
                                    & \textit{Model 4 (M4):} Input-8-4-3-output\\
                                    & \textit{Model 5 (M5):} Input-16-8-output\\
                                    & \textit{Model 6 (M6):} Input-2-4-8-output\\
\bottomrule
\end{tabular}%
}
\end{table}

\subsection{Results}
\subsubsection{Overall Performance Cost}

Figure \ref{fig:cost} illustrates the fault-tolerance cost across small, medium, and large scales under incremental drift. Before the drift (indicated by the vertical dashed line), all algorithms showed comparable performance under stationary conditions. However, the introduction of drift triggered a clear divergence in the results. The CND, BP-CND, and ES-CND exhibited an increase in cost. In contrast, the EES-CND maintained the lowest cost and showed rapid stabilisation after the drift. These results confirm that the EES-CND effectively mitigates cumulative performance loss to provide enhanced stability and long-term efficiency in non-stationary and failure-prone systems.

\subsubsection{Service Recovery Time}
\noindent\textbf{Reconfiguration Time:}
Figure \ref{fig:reconf} illustrates the service reconfiguration time. Before the drift, all the algorithms showed comparable performance. After the drift began, a clear divergence emerged. The CND, BP-CND, and ES-CND experienced significant increases in reconfiguration time, whereas the EES-CND maintained the lowest value and was more stable.

\noindent\textbf{Computational Overhead:} Table \ref{tab:runtime} details the runtime per failure interval across all system scales. The CND exhibits the lowest overhead because it lacks an adaptive layer. BP-CND benefits from the low computational cost of single-sample backpropagation updates compared with that of evolutionary approaches. ES-CND and EES-CND incur higher computational overheads owing to perturbation sampling and multiple evaluations. However, their runtimes remain modest, reaching only $0.71$ s and $0.78$ s, respectively, in large-scale scenarios. This shows that even with a large number of failures (e.g., over 60 simultaneous component failures), the runtime remains below 1 s.

\begin{table}
\centering
\caption{Algorithm runtime per failure interval (s)}
\label{tab:runtime}
\resizebox{0.98\linewidth}{!}{%
\begin{tabular}{lccc}
\toprule
\textbf{Method} & \textbf{Small-scale} & \textbf{Medium-scale} & \textbf{Large-scale} \\
\toprule
EES-CND   & 0.05  & 0.18  & 0.78 \\
\midrule
ES-CND    & 0.05  & 0.18  & 0.71 \\
\midrule
BP-CND    & 0.03  & 0.09  & 0.49  \\
\midrule
CND       & 0.01   & 0.02  & 0.26  \\
\bottomrule
\end{tabular}%
}
\end{table}

\noindent\textbf{Downtime:}
Table \ref{tab:recovery} summarises the average service recovery times per failure interval across all scales. Despite the slightly higher runtime noted in Table \ref{tab:runtime}, EES-CND achieved the lowest overall recovery time (downtime) in every scenario. Although CND and BP-CND are computationally faster, they yield inferior placement strategies that increase reconfiguration delays. Thus, the modest computational overhead of EES-CND is offset by a reduction in downtime, resulting in superior end-to-end performance.

\begin{table}
\centering
\caption{Downtime per failure interval (s)}
\label{tab:recovery}
\resizebox{0.98\linewidth}{!}{%
\begin{tabular}{lccc}
\toprule
\textbf{Method} & \textbf{Small-scale} & \textbf{Medium-scale} & \textbf{Large-scale} \\
\toprule
EES-CND   & 3.19  & 3.60  & 4.32 \\
\midrule
ES-CND    & 3.45  & 3.89  & 4.52 \\
\midrule
BP-CND    & 3.50  & 4  & 4.48  \\
\midrule
CND       & 3.78   & 4.1  & 4.49  \\
\bottomrule
\end{tabular}%
}
\end{table}

\subsubsection{Service Response Time}
\begin{figure*}
    \centering
    \includegraphics[width=0.97\linewidth]{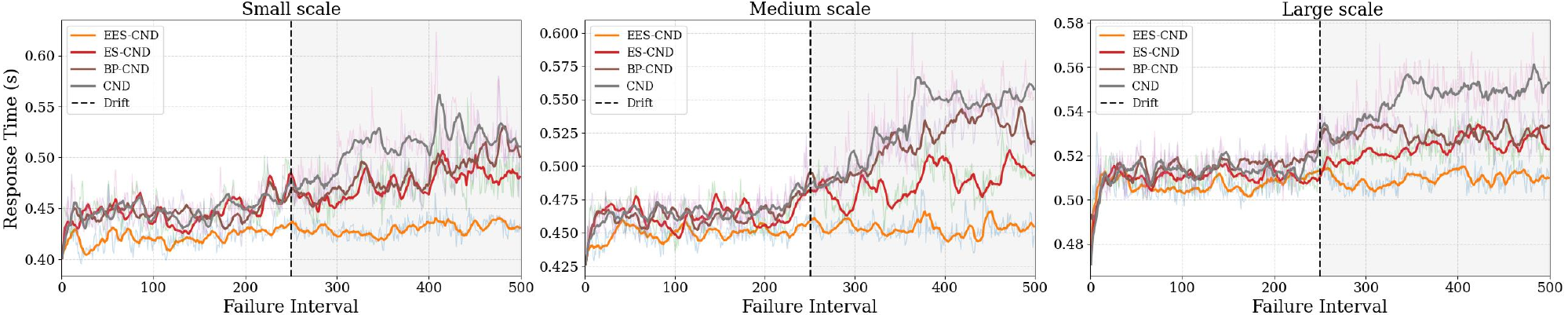}
    \caption{Service response time after recovery (pale lines: actual data; plain lines: 10-point moving average).}
    \label{fig:restime}
\end{figure*}

As shown in Figure \ref{fig:restime}, EES-CND achieved the lowest response time across all scales, indicating that its reconfigurations are better aligned with evolving system conditions. After the drift, the response times of the other techniques shifted upward, as their adaptation mechanisms failed to track the changing system correlations. In contrast, EES-CND maintained a stable response time throughout the drift phase. These results demonstrate that EES-CND preserves the post-recovery QoS, which is critical for real-time failure-prone edge-cloud systems.

\begin{figure*}
    \centering
    \includegraphics[width=0.97\linewidth]{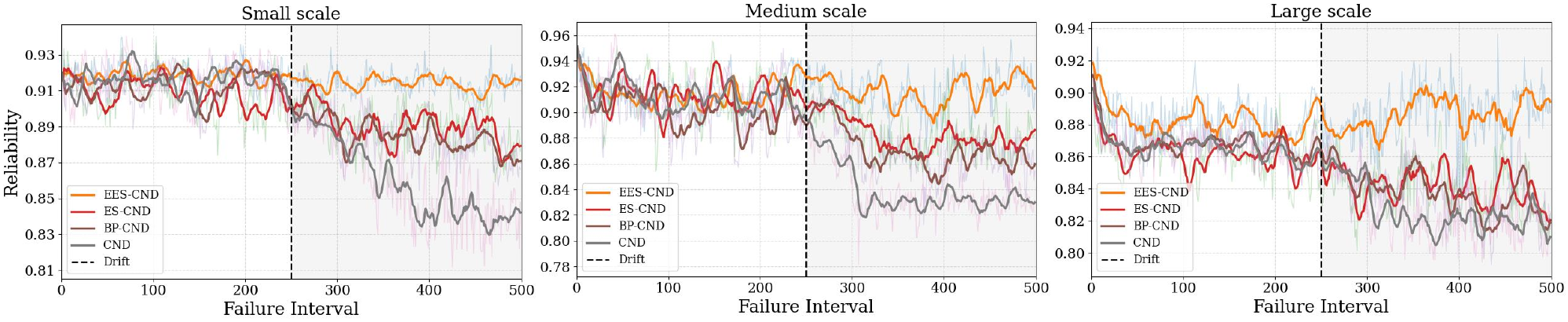}
    \caption{Service reliability after recovery (pale lines: actual data; plain lines: 10-point moving average).}
    \label{fig:reliability}
\end{figure*}

\subsubsection{Service Reliability}
Figure \ref{fig:reliability} shows that once the incremental drift begins, the reliability of the CND, BP-CND, and ES-CND exhibits a continuous downward trend across all scales. In contrast, the EES-CND demonstrates strong drift resilience, maintaining the highest post-recovery reliability with only moderate fluctuations. By maintaining high reliability, the EES-CND ensures long-term service stability and significantly mitigates the risk of cascading failures.

\subsubsection{Impact of Model Collaboration }
\begin{figure}
    
    \includegraphics[width=0.95\linewidth]{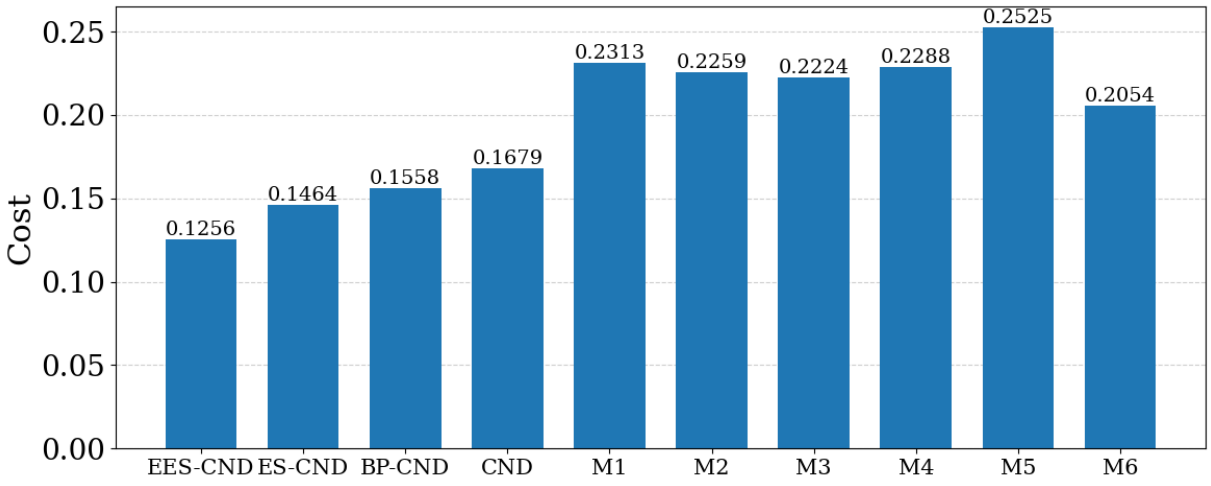}
    \caption{EES-CND vs individual models cost}
    \label{fig:models}
\end{figure}

This section compares EES-CND with its individual models by evaluating each one independently under the same experimental conditions. To maintain conciseness and avoid redundancy, we focused our analysis on large-scale scenarios because similar trends were observed across all scales. Figure \ref{fig:models} shows the average fault tolerance cost after 500 failure events for collaborative mechanisms versus individual models (M1--M6). EES-CND achieved the lowest cost, significantly outperforming all alternatives. Standalone models had much higher costs. On average, EES-CND reduced cost by $44.8\%$, showing that its collaborative strategy uses complementary decision patterns compared to single models.

\subsubsection{Impact of Collaboration Size}
\begin{figure}
    \centering
    \includegraphics[width=0.9\linewidth]{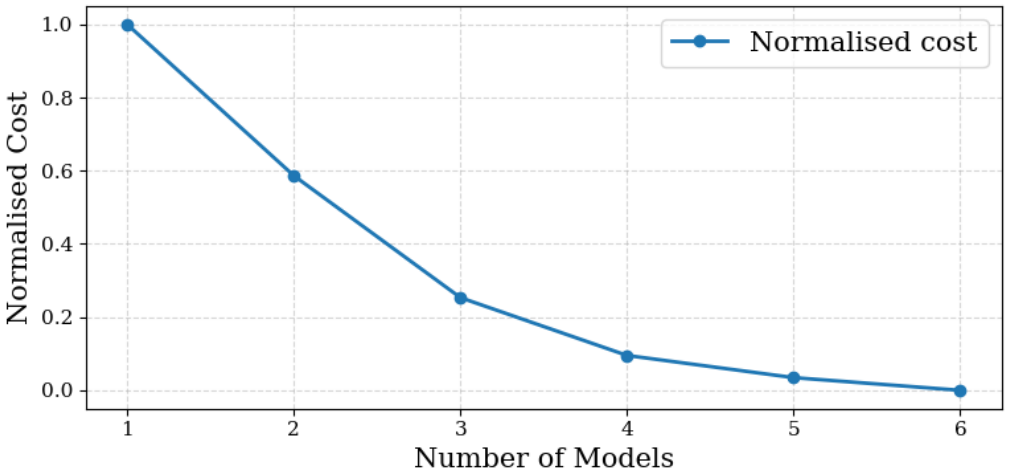}
    \caption{Cost versus number of models}
    \label{fig:numModels}
\end{figure}

The effectiveness of collaborative decision-making depends on the number of models. Although adding more models improves decision quality through complementary knowledge, it also increases computational cost and redundancy. As shown in Figure \ref{fig:numModels}, the performance improves significantly from a single model, but the gains diminish as more models are added. Based on this empirical analysis, we set the number of collaborative models to six.

\section{\uppercase{Conclusion}}\label{conclusion}
This study formulates fault-tolerant service placement in dynamic edge–cloud environments as an online adaptive optimisation problem under multilevel drift. To mitigate performance degradation, we propose EES-CND, which integrates pretrained models with adaptive models updated using an enhanced evolution strategy. The simulation results show that EES-CND reduces the cost by 44.8\% compared with standalone models and maintains stable post-recovery response times and reliability. These findings confirm that the combination of collaborative decision-making with adaptive evolutionary updates ensures long-term stability in non-stationary environments. Future research will explore hybrid proactive–reactive strategies to further enhance the resilience of large-scale edge-cloud infrastructures.
\section*{\uppercase{Acknowledgements}}
Parts of this study were supported by the Knowledge Foundation of Sweden (KKS).

\bibliographystyle{apalike}
{\small
\bibliography{ref}}

\end{document}